\documentclass[journal,transmag]{IEEEtran}
\usepackage{cite}

\ifCLASSINFOpdf
\usepackage[pdftex]{graphicx}
\graphicspath{figures/}
\DeclareGraphicsExtensions{.pdf,.jpeg,.png}
\else
\usepackage[dvips]{graphicx}
\graphicspath{figures/}
\DeclareGraphicsExtensions{.eps}
\fi
\usepackage{amsmath}
\usepackage{algorithm}
\usepackage{algorithmic}

\usepackage{array}

\ifCLASSOPTIONcompsoc
\usepackage[caption=false,font=normalsize,labelfont=sf,textfont=sf]{subfig}
\else
\usepackage[caption=false,font=footnotesize]{subfig}
\fi
	\usepackage{url}
	\usepackage{hyperref}
	\hypersetup{%
		unicode=true,      
		pdffitwindow=true, 
		pdfnewwindow=true, 
		pdfkeywords={},    
		colorlinks=true,   
		linkcolor=black,   
		citecolor=black,   
		filecolor=black,   
		urlcolor=black,    
		pdftitle={An Efficient Architecture and High-Throughput Implementation of CCSDS-123.0-B-2 Hybrid Entropy Coder Targeting Space-Grade SRAM FPGA Technology},       
		pdfauthor={P. Chatziantoniou, A. Tsigkanos, D. Theodoropoulos, N. Kranitis, A. Paschalis},      
		pdfsubject={}      
	}
	
	\usepackage{caption}
	\usepackage{todonotes}
	\usepackage{booktabs}
	\usepackage{multirow}
	\usepackage{placeins} 
	\usepackage{easyReview}

	\newcommand\copyrighttext{%
		\footnotesize \textcopyright 2022 IEEE. Personal use of this material is permitted.
		Permission from IEEE must be obtained for all other uses, in any current or future 
		media, including reprinting/republishing this material for advertising or promotional 
		purposes, creating new collective works, for resale or redistribution to servers or 
		lists, or reuse of any copyrighted component of this work in other works. 
		DOI: \href{https://dx.doi.org/10.1109/TAES.2022.3173583}{10.1109/TAES.2022.3173583}}
	\newcommand\copyrightnotice{%
		\begin{tikzpicture}[remember picture,overlay]
			\node[anchor=south,yshift=10pt] at (current page.south) {\fbox{\parbox{\dimexpr\textwidth-\fboxsep-\fboxrule\relax}{\copyrighttext}}};
		\end{tikzpicture}%
	}

\begin{document}
		%
		\title{An Efficient Architecture and High-Throughput Implementation of CCSDS-123.0-B-2 Hybrid Entropy Coder Targeting Space-Grade SRAM FPGA Technology}

		
		\author{\IEEEauthorblockN{Panagiotis Chatziantoniou\IEEEauthorrefmark{1},
				Antonis Tsigkanos\IEEEauthorrefmark{1},
				Dimitris Theodoropoulos\IEEEauthorrefmark{1},
				Nektarios Kranitis\IEEEauthorrefmark{2} and
				Antonios Paschalis\IEEEauthorrefmark{1}}
			\IEEEauthorblockA{\IEEEauthorrefmark{1}Dept. of Informatics \& Telecommunications, National and Kapodistrian University of Athens\\
			}
			\IEEEauthorblockA{\IEEEauthorrefmark{2}Dept. of Aerospace Science \& Technology, National and Kapodistrian University of Athens}
			}
		
		%

		
		\IEEEtitleabstractindextext{%
			\begin{abstract}
				Nowadays, hyperspectral imaging is recognized as a cornerstone remote sensing technology. The explosive growth in image data volume and instrument data rates, compete with limited on-board storage resources and downlink bandwidth, making hyperspectral image data compression a mission critical on-board processing task. Recently, the Consultative Committee for Space Data Systems (CCSDS) extended the previous issue of the CCSDS-123.0 Recommended Standard for multi- and hyperspectral image compression to provide with Near-Lossless compression functionality. A key feature of the CCSDS-123.0-B-2 is the improved Hybrid Entropy Coder, which at low bit rates, provides substantially better compression performance than the Issue 1 entropy coders. In this paper, we introduce a high-throughput hardware implementation of the CCSDS-123.0-B-2 Hybrid Entropy Coder. The introduced architecture exploits the systolic design pattern to provide modularity and latency insensitivity in a deep and elastic pipeline achieving a constant throughput of 1 sample/cycle with a small FPGA resource footprint. This architecture is described in portable VHDL RTL and it is implemented, validated and demonstrated on a commercially available Xilinx KCU105 development board hosting a Xilinx Kintex Ultrascale XCKU040 SRAM FPGA, and thus, is directly transferable to the Xilinx Radiation Tolerant Kintex UltraScale XQRKU060 space-grade devices for space deployments. Moreover, state-of-the-art SpaceFibre (ECSS-E-ST-50-11C) serial link interface and test equipment were used in the validation platform to emulate an on-board deployment. The introduced CCSDS-123.0-B-2 Hybrid Entropy Encoder achieves a constant throughput performance of 305 MSamples/s. To the best of our knowledge, this is the first published fully-compliant architecture and high-throughput implementation of the CCSDS-123.0-B-2 Hybrid Entropy Coder, targeting space-grade FPGA technology.
			\end{abstract}
			
			}

		\maketitle
		
		\copyrightnotice
		
		\IEEEdisplaynontitleabstractindextext

		%
		\IEEEpeerreviewmaketitle

		\section{\textbf{Introduction}}
		\IEEEPARstart{H}{yperspectral} imaging is recognized as a cornerstone remote sensing technology. The latest
		high-resolution and high-speed space-borne imagers have brought an explosive growth in data volume.
		For example, the HyspIRI sensor developed by NASA can produce up to 5 TB of data per day.
		This competes with the limited on-board storage resources and downlink bandwidth, making hyperspectral image compression a mission critical on-board processing task.
		Due to the high data volume reduction often needed to meet spacecraft downlink bandwidth requirements, lossy compression is becoming increasingly
		important. In this context, the Multispectral Hyperspectral Data Compression (SLS-MHDC) Working Group of the
		Consultative Committee for Space Data Systems (CCSDS) standardized the new Issue 2 ``Low-Complexity Lossless and
		Near-Lossless Multispectral and Hyperspectral Image Compression'' standard CCSDS-123.0-B-2~\cite{ccsds123_b2_blue_book}.
		This new Issue 2 extends Issue 1~\cite{ccsds123_b1_blue_book}, incorporating support for low-complexity near-lossless compression,
		while retaining lossless
		compression capabilities. Near-lossless refers to the ability to perform compression in a way that limits the maximum error in the reconstructed image to a user-specified bound.
		
		A key feature of CCSDS-123.0-B-2 is the improved Hybrid Entropy Coder option. At high bit-rates, the Hybrid Entropy Coder
		encodes most samples using a family of codes that are equivalent to those used by the Sample-Adaptive Encoder of Issue
		1, and thus, has nearly identical high-bit-rate performance. However, at low bit rates it has substantially better performance than
		the Issue 1 entropy encoders~\cite[p.~4-29]{ccsds123_green_book}. For example, the Sample-Adaptive Encoder of Issue 1 cannot reach bit-rates 
		lower than 1 bit-per-sample due to design constraints, while the Rice-based Block-Adaptive Encoder (described in CCSDS-121.0-B-3) may, but at a non-negligible bit-rate overhead.
		
		The Hybrid Entropy Coder specified in CCSDS-123.0-B-2 is an extended version of the NASA FLEX original hybrid entropy
		coder~\cite{Klimesh_flex_2005},\cite{Klimesh_flex_2006} so that decoding proceeds in reverse order. This permits a more memory-efficient
		implementation than FLEX’s original entropy coder, which was based on an interleaved entropy coding approach.
		The Hybrid Entropy Coder includes codes equivalent to the
		Length-Limited Golomb-Power-of-2 codes used by the Sample-Adaptive Entropy Coder with the addition of 16
		variable-to-variable length “low-entropy” codes to provide better compression of low-entropy data. Such low-entropy
		data become more prevalent as increased predictor quantization step sizes are used i.e. increasingly lossy compression. The Hybrid Entropy Coder
		adaptively switches between high and low entropy encoding methods on a sample-by-sample basis, using code selection
		statistics similar to those used by the Sample-Adaptive coder. A single output codeword from a low-entropy code may
		encode multiple samples, which allows obtaining lower compressed data rates than can be produced by the Sample-Adaptive
		Entropy Coder.
		
		Apart from the compression needs, on-board applications require devices that are capable of high-performance with low power consumption
		and radiation hardness characteristics. The current state-of-the-art SRAM-based FPGA technology offers radiation hardening by design
		(RHBD), high density and dynamic partial reconfiguration for in-flight adaptability and Time-Space Partitioning (TSP)
		of on-board data processing tasks.
		An excellent example of such technology is the RHBD Xilinx Kintex-Ultrascale XQRKU060 FPGA which provides exceptional hardness
		to Single-Event-Upset (SEU), typical immunity of 80MeV-cm$^2$/mg to Single-Event Latchup (SEL), data path protection from Single-Event Transients
		(SET) and maximum tolerance of 100 Krad to Total Ionizing Dose (TID)~\cite{kintex_rad_hard}.
		The XQRKU060 FPGA offers those technological advantages and is considered a suitable device for on-board payload data
		processing applications due to its ability to support upgrades after launch, greatly enhancing mission profile and extending valuable mission life time.
		
		The first, fully-compliant, architecture and implementation of CCSDS-123.0-B-2 Hybrid Entropy Coder was presented in
		\cite{chatziant_obpdc2020}.	
		The architecture achieved variable throughput performance depending on hyperspectral image
		statistics operating at 1 sample/cycle only for high-entropy data and at no less than 0.33 samples-per-cycle, for low-entropy data.  
		The maximum throughput (1 sample/cycle) which was achieved for lossless compression configuration of a high entropy
		hyperspectral cube was 344 MSamples/sec targeting the XQRKU060 space-grade FPGA. 
		However, even for lossless compression of low-entropy data or near-lossless mode, where low-entropy coded samples occur increasingly more often, the throughput
		performance is degraded when the Absolute Error Limit Constant increases, with a lower bound of 114 MSample/sec.
		
		In this paper, we introduce an efficient architecture and high-throughput hardware implementation of the CCSDS-123.0-B-2 Hybrid Entropy
		Coder. 
		The proposed architecture extends our previous work in \cite{chatziant_obpdc2020} achieving a constant throughput of 1
		sample/cycle introducing an efficient codetable lookup by modifying the low entropy coder without any performance degradation when using near-lossless mode even with larger values of error limits.	
		Moreover, the proposed
		architecture is implemented in portable VHDL RTL and exploits the systolic design
		pattern to provide modularity and latency insensitivity in a deep and elastic pipeline  
		minimizing the number of stalls.
		The introduced Hybrid Entropy
		Coder architecture is validated and demonstrated on a commercially available Xilinx KCU105 development board hosting a
		Xilinx Kintex Ultrascale XCKU040 SRAM FPGA, and is therefore directly transferable to the Xilinx Radiation Tolerant Kintex
		UltraScale XQRKU060 space-grade devices for space deployments. Moreover, state-of-the-art SpaceFibre (ECSS-E-ST-50-11C)
		high-speed serial link interface and test equipment were used in the validation platform to match space deployment.
		The introduced CCSDS-123.0-B-2 Hybrid Entropy Encoder achieves a constant high-throughput performance of 305 MSamples/s
		(4.88 Gbps @ 16bpppb), with minimal footprint that is 2.10\% (5086) of device LUTs and 0.17\% (1) BRAMs of FPGA resources.
		To the best of our knowledge, this is the first published, fully-compliant, architecture and high-throughput implementation of the
		CCSDS-123.0-B-2 Hybrid Entropy Coder, also targeting space-grade FPGA technology.
		
		The rest of this contribution is organized as follows: Section 2 provides background information about the CCSDS-123.0-B-2
		Recommended Standard and the Hybrid Entropy Coder algorithm, while Section 3 describes the introduced architecture. Section
		4 provides experimental results including the verification of the proposed architecture, the validation of the implemented
		design on Xilinx KCU105 development board interfacing with SpaceFibre, as
		well as resource and throughput performance statistics of the implemented design. Section 5 presents related work and comparisons. Finally, Section 6 concludes the paper.
		
		\begin{figure}[t]
			\centering
			\captionsetup{justification=centering}
			\includegraphics[width=\linewidth]{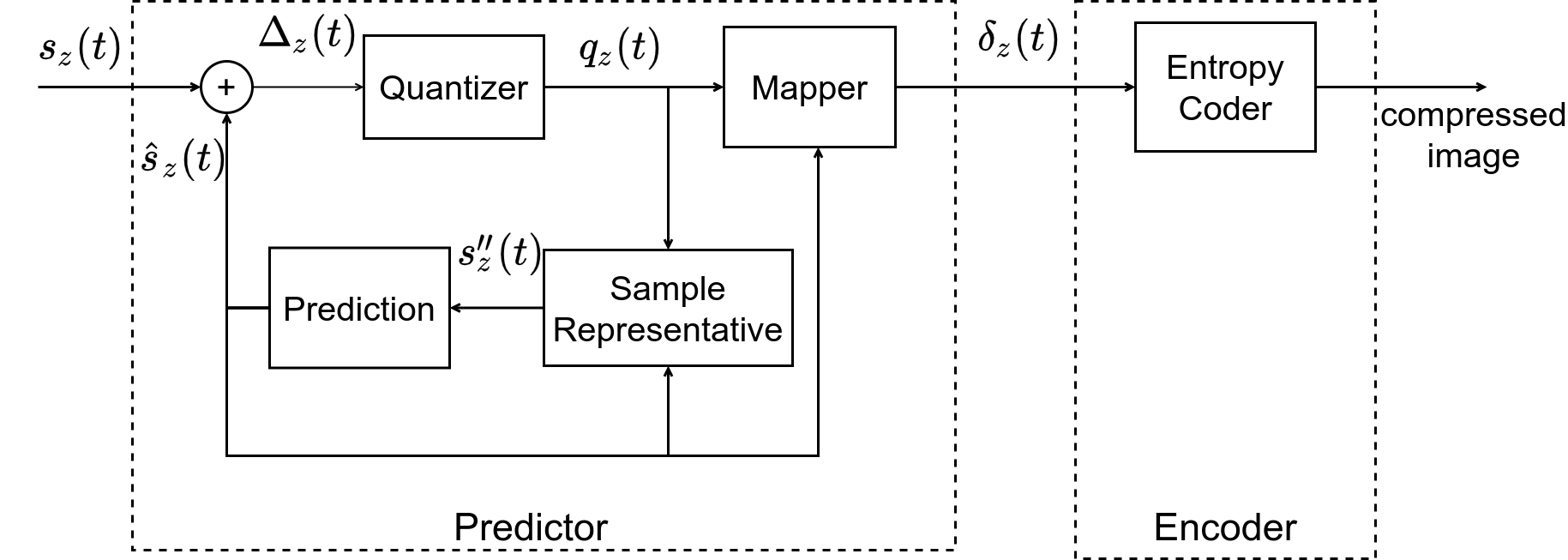}
			\caption{Block diagram of the CCSDS-123.0-B-2 compressor~\cite{ccsds123_b2_blue_book}}
			\label{fig:compressor}
		\end{figure}
		
		\section{\textbf{Background}}
		\subsection{\textbf{CCSDS-123.0-B-2 Overview}}
		The CCSDS-123.0-B-2 standard was designed to provide an effective method of performing lossless or near-lossless compression of three-dimensional
		image data with low implementation complexity for space-borne imagers. Near-lossless compression refers to the ability to perform compression such
		that the maximum error in the reconstructed image can be limited to a user-specified bound by adjusting the absolute and relative error parameters.
		
		Incoming image samples enter at compressor's input. Image indices are denoted as $s_z(t)$ where $t=y\cdot N_x+x$, $x=0,...,N_x-1$, $y=0,...,N_y-1$
		are the spacial coordinates ($N_x$ columns and $N_y$ rows) and $z=0,...,N_z-1$ the spectral dimension. Image samples produced by multispectral and hyperspectral imagers are typically interleaved in one of three common orderings:
		z,y,x (Band SeQuential [BSQ]),
		y,x,z (Band Interleaved Pixels [BIP]),
		and y,z,x (Band Interleaved Lines [BIL]).
		In BSQ the compression of all image samples of a spectral band is computed
		before processing the following bands; 
		in BIP a sample is compressed for all the bands before processing next samples; 
		finally, in BIL each line of samples is compressed for all the bands before processing the next lines.
		
		The predictor uses a low-complexity adaptive linear prediction method to predict the value of
		each sample based on the values of nearby samples in a small three-dimensional
		neighborhood. Prediction can be performed causally in a single pass through the image, making use of an
		adaptively weighted prediction algorithm. Since the original input samples will not be available to the decompressor due to lossy compression, the
		predictor performs calculations based on sample representatives $s''_z(t)$ instead.
		
		Besides using sample representatives, the predictor in Issue 2 also differs from Issue 1 in that each prediction residual $\Delta_z(t)$, that is, the difference between the
		predicted and actual sample values, is quantized using a uniform quantizer. The quantizer step size can be
		controlled via an absolute error limit (so that samples can be reconstructed with a user-specified error bound) and/or a relative error limit (so that
		samples predicted to have smaller magnitude can be reconstructed with lower error). Lossless compression in a band is
		obtained by setting the absolute error limit to zero. The quantized prediction residual $q_z(t)$ is then mapped to an unsigned integer mapped
		quantizer index $\delta_z(t)$. These mapped quantizer indices make up the output of the predictor.
		
		The Encoder receives those mapped quantizer indices from the Predictor and encodes them using a family of codes. The
		Standard describes three possible Encoder options, the Sample Adaptive Encoder, Block-Adaptive Encoder and Hybrid Entropy Coder. The CCSDS-120.2-G-2 Informational Report~\cite{ccsds123_green_book} includes detailed
		benchmarks of the encoders and highlights that even though the Hybrid Entropy Coder is the more complex encoder option,
		it is capable of improved compression performance for both lossless and near-lossless compression for well-chosen parameters. A comprehensive review of the Standard can also be found
		at~\cite{hernandez-cabronero_ccsds_2021}.
		
		\subsection{\textbf{Hybrid Entropy Encoding Algorithm}}
		The Hybrid Entropy Coder is a modified version of the one originally used by the FLEX entropy coder. It includes codes
		equivalent to the length-limited Golomb Power-of-2 (GPO2) codes (i.e. Golomb codes with parameters that are powers of 2, also known as Golomb-Rice codes~\cite[p.~3-15]{ccsds123_green_book}) used by the Sample-Adaptive encoder, but extended with
		an additional 16 variable-to-variable length low-entropy codes. During encoding it adaptively switches between
		these two coding methods on a sample-by-sample basis based on code selection statistics. A single output codeword from a low-entropy code may encode
		multiple samples, which allows obtaining lower compression data rates (under one bit-per-pixel) than those achievable by the
		Sample-Adaptive entropy coder.
		
		The mapped quantizer indices, $\delta_z(t)$ of dynamic range $D$, from the Predictor's output are the inputs of the encoder.
		The Coder maintains the Adaptive Code Selection Statistics (ACSS), a high-resolution accumulator,
		$\tilde{\Sigma}_z(t)$, and a counter $\Gamma(t)$. Based on the ratio of these variables, the running
		$\delta_z(t)$ is encoded with a high entropy or a low entropy code.
		
		Initially when $t=0$, both variables are initialized, and the first sample of every band, $\delta_z(0)$, is emitted uncompressed.
		For the rest of the encoding process, both variables are updated before coding the sample and rescaled when the counter saturates (indicated by the rescaling factor $\gamma^{*}$), as
		shown in~(\ref{eq: sigma_rescale}) and~(\ref{eq: gamma_rescale}).
		After rescaling, the most significant value of $\tilde{\Sigma}_z(t)$ is emitted to enable recalculation of the accumulator during the decoding process.
		
		\begin{algorithm}[t]
			\caption{Functionality of Hybrid Entropy Coder}
			\label{lst:HEC_psedocode}
			\begin{algorithmic}
				\IF{$t = 0$}
				\STATE init()
				\ELSIF{$t = N_xN_y-1$}
				\STATE compressed\_image\_tail()
				\ELSE
				\STATE \big($\Sigma_z(t),\Gamma(t)$\big) $\leftarrow$ update\_acss\big($\Sigma_z(t-1),\Gamma(t-1),\delta_z(t)$\big)
				
				\STATE hilo $\leftarrow$ entropy\_coder\_selection\big($\Sigma_z(t),\Gamma(t)$\big)
				
				\IF{hilo $= 1$}
				\STATE codeword $\leftarrow$ high\_entropy\_coder\big($\delta_z(t)$\big)
				\STATE to\_bitstream(codeword)
				\ELSE
				\STATE codeword $\leftarrow$ low\_entropy\_coder\big($\delta_z(t)$\big)
				\IF{codeword\_match $= 1$}	
				\STATE to\_bitstream(codeword)
				\ENDIF
				\ENDIF
				\ENDIF
			\end{algorithmic}
		\end{algorithm}	
		
		\begin{footnotesize}
			\begin{equation}
				\tilde{\Sigma}_{z}(t)=\left\{
				\begin{array}{ll}
					\tilde{\Sigma}_{z}(t-1)+4 \delta_{z}(t) & ,\Gamma(t-1)<2^{\gamma^{*}}-1 \\
					{\Bigg\lfloor\dfrac{\tilde{\Sigma}_{z}(t-1)+4 \delta_{z}(t)+1}{2}\Bigg\rfloor} & ,\Gamma(t-1)=2^{\gamma^{*}}-1
				\end{array}\right.
				\label{eq: sigma_rescale}
			\end{equation}
		\end{footnotesize}
		
		\begin{small}
			\begin{equation}
				\Gamma(t)=\left\{
				\begin{array}{ll}
					\Gamma(t-1)+1 & ,\Gamma(t-1)<2^{\gamma^{*}}-1 \\
					{\Bigg\lfloor\dfrac{\Gamma(t-1)+1}{2}\Bigg\rfloor} & ,\Gamma(t-1)=2^{\gamma^{*}}-1
				\end{array}\right.
				\label{eq: gamma_rescale}
			\end{equation}
		\end{small}
		
		\begin{figure*}[!ht]
			\centering
			\captionsetup{justification=centering}
			\includegraphics[width=0.95\linewidth]{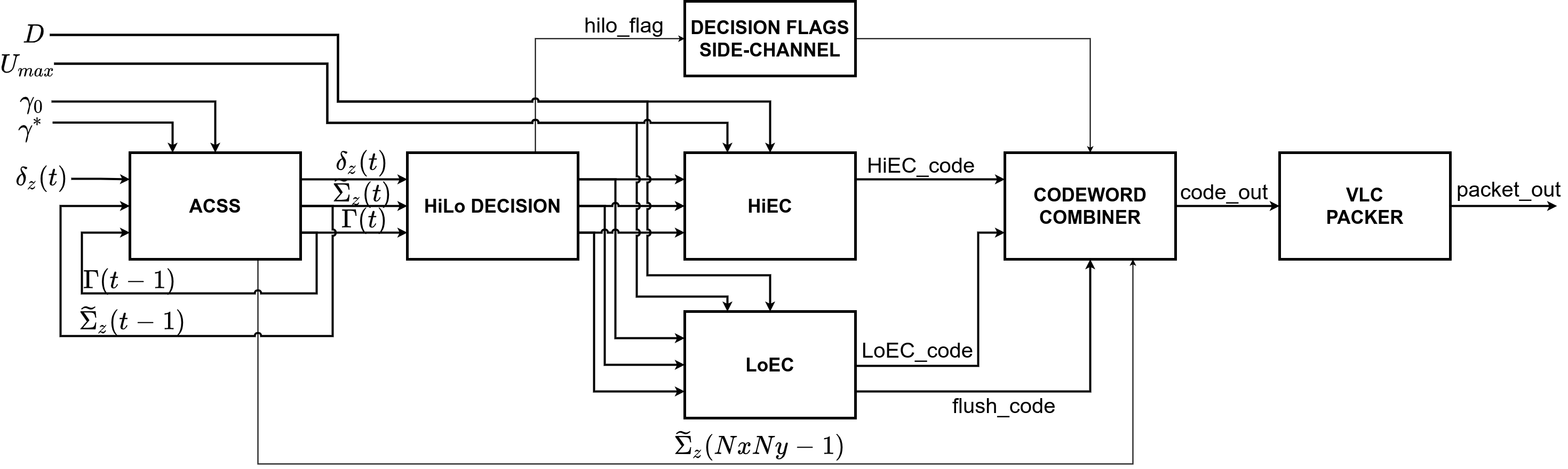}
			\caption{Top-level architecture for the proposed Hybrid Entropy Coder}
			\label{fig:top_level}
		\end{figure*}
		
		In equation~(\ref{eq: hilo_sel}), the choice of coder is represented by the high/low entropy flag ($hilo$), where $T_0$ is a constant provided by the
		standard. When the flag is set, the current sample shall be encoded with a high entropy code, otherwise with a low entropy code.
		\begin{equation}
			hilo=\left\{
			\begin{array}{ll}
				1 & ,\tilde{\Sigma}_z(t)\cdot2^{14} \leq T_0\cdot\Gamma(t) \\
				0 & ,else
			\end{array}\right.
			\label{eq: hilo_sel}
		\end{equation}
		
		Each high entropy sample is then encoded by a Reverse Length-Limited Golomb Power-of-2 (RLL-GPO2) code. Each code is identified by a code index
		$k_z(t)$, being the largest positive integer satisfying
		\begin{equation}
			kz(t) \leq max(D-2, 2)
			\label{eq: kz_limit}
		\end{equation}
		\begin{equation}
			\Gamma(t)2^{k_z(t)} \leq \tilde{\Sigma}_z(t) + \bigg\lfloor\dfrac{49}{2^5}\Gamma(t)\bigg\rfloor
			\label{eq: kz_high}
		\end{equation}
		
		The RLL-GPO2 codeword for the high entropy sample $\delta_z(t)$, $\Re_{k_z(t)}^{\prime}(\delta_z(t))$ is defined as follows:
		\begin{enumerate}
			\item[a)] if $\lfloor \delta_z(t)/2^{k_z(t)}\rfloor < U_{max}$, where $U_{max}$ is the maximum unary length, then
			$\Re_{k_z(t)}^{\prime}(\delta_z(t))$
			consists of the $k_z(t)$ least significant
			bits of the binary representation of $\delta_z(t)$, followed by a ‘one’, followed by $\lfloor  \delta_z(t) / 2^k_z(t)\rfloor$ ‘zeros’
			\item[b)] otherwise, $\Re_{k}^{\prime}(\delta_z(t))$ consists of the $D$-bit binary representation of $\delta_z(t)$ followed by
			$U_{max}$ ‘zeros’.
		\end{enumerate}
		
		Low entropy samples are encoded using one of 16 variable-to-variable length family of codes. The code index of the low entropy code to be used for encoding the low entropy sample $\delta_z(t)$ is the largest positive $i$ satisfying
		\begin{equation}
			\tilde{\Sigma}_z(t)\cdot2^{14} \leq T_i\cdot\Gamma(t),\ 0\leq i\leq 15
			\label{eq: index_sel}
		\end{equation}
		where $T_0,...,T_{15}$ are constants provided by the standard and $T_0$ is used in~(\ref{eq: kz_high}).
		
		For each code $i$ a prefix of previously input samples is maintained. When a sample is processed, a symbol is added to the corresponding prefix. The
		standard defines a list of complete prefixes for each code. When a code's prefix matches a complete prefix, then a unique codeword
		corresponding to that sequence of symbols is emitted and the prefix is cleared for this code $i$.
		

		The complete prefixes of any low entropy code $i$, can only contain samples satisfying $\delta_z(t) \leq L_i$, where $L_0,...,L_{15}$ are constant symbol
		limits provided by the standard. When a sample exceeds this limitation, then $\Re_{0}(\delta_z(t)-L_i-1)$ is emitted, and an escape symbol $X$ (here represented as $L_i+1$) is added to
		the code's prefix. The addition of the escape symbol, completes the prefix and the corresponding codeword is emitted. Therefore in this case an RLL-GPO2
		codeword is followed by a low entropy codeword.
		The input symbol selection can be summarized as
		\begin{equation}
			\iota_{z}(t)=\left\{
			\begin{array}{cl}
				\delta_{z}(t) & ,\delta_{z}(t) \leq L_{i} \\
				L_i + 1 & ,\delta_{z}(t)>L_{i}
			\end{array}\right.
			\label{eq: sym_calc}
		\end{equation}
		
		A more detailed description of the encoding procedure including code tables and limits, can be found in the Standard
		\cite{ccsds123_b2_blue_book}. The encoding procedure is outlined in pseudocode in Algorithm \ref{lst:HEC_psedocode}.
		
		\section{\textbf{Introduced Architecture}}
		The Hybrid Entropy Coder is designed as an IP core described in technology agnostic VHDL RTL. However, technology specific blocks (e.g. DSP48E blocks in Xilinx FPGAs) are used by inference, as well as, generic usage for memory technology mapping between inference and vendor specific memory cells.
		The encoder operates in the BIP ordering to be matched with a lossless predictor in BIP order (e.g~\cite{tsigkanos_123b1_2018}), however, it can be modified to match specific instrument sensors and mission requirements, with
		different pixel order (BIL and BSQ). The top-level block diagram of the proposed Hybrid Entropy Coder architecture implemented as an IP core is shown in~\autoref{fig:top_level}.
		
		The IP Core interfaces using AXI4-Stream based I/O, supporting flow control using the protocol handshake (\texttt{tvalid}
		and \texttt{tready} signals). Internally, each encoder sub-unit, as shown in~\autoref{fig:top_level}, is pipelined using a systolic latency insensitive design
		pattern, with elastic buffers~\cite{ebs} as pipeline registers. The elastic buffers use AXI4-Stream
		handshaking and allow for full throughput (1 cycle/sample) when neither source or sink are stalling. This
		design pattern avoids additional controllers for flow control, or superfluous buffering to manage sink side stalls. At the same time it
		facilitates Unit testing, by having consistent, verified interfaces in internal components on which testbench bus functional models attach
		with a consistent protocol.
		
		The Hybrid Entropy Coder comprises of six pipelined components at the top level, which are:
		\begin{enumerate}
			\item \textit{Adaptive Code Selection Statistics (ACSS) Unit}
			\item \textit{High/Low (HiLo) Entropy Decision Unit}
			\item \textit{High Entropy Coder (HiEC) Unit}
			\item \textit{Low Entropy Coder (LoEC) Unit}
			\item \textit{Codeword Combiner Unit} and
			\item \textit{Variable Length Code (VLC) Packer Unit}
		\end{enumerate}
		
		\subsection{\textbf{Design Considerations for Hardware Implementation}}
		Using the latency insensitive design pattern, feed-forward processing paths are further pipelined to decrease
		logic path lengths and increase achievable frequency. However, processing feedback loops imply a total number of delay cycles equal to the pipeline registers in the loop, which if exceeded, limits performance in terms of cycles/sample processed. This
		creates a pipeline depth versus critical path (achievable F\textsubscript{max}) trade-off to be considered in these
		feedback loops. In this context, two components stand out in complexity, the \texttt{ACSS} unit, and the \texttt{LoEC}
		unit, which contain such feedback loops.
		
		The \texttt{ACSS} unit contains a feedback loop in the update of $\tilde{\Sigma}_z(t)$ and $\Gamma(t)$. Initializing
		and computing this update depending on previously computed values ($\tilde{\Sigma}_z(t-1)$, $\Gamma(t-1)$), is handled
		by a \textit{Loop Controller} module. In this unit, for BIP and BIL order the feedback datapath delay is commonly larger than the pipeline depth
		($N_z$ clock cycles), therefore the loop does not cause a performance degradation in terms of samples-per-cycle processed, unless the number of bands is extremely small. For BSQ order, the feedback datapath comprises of exactly one clock cycle delay, regardless of the number of bands.
		
		The \texttt{LoEC} unit contains a loop where the input codeword in a code-table lookup operation, depends on the output
		of previous lookups of the same code-table. 
				
		\begin{figure}[!t]
			\centering
			\captionsetup{justification=centering}
			\includegraphics[width=0.95\linewidth]{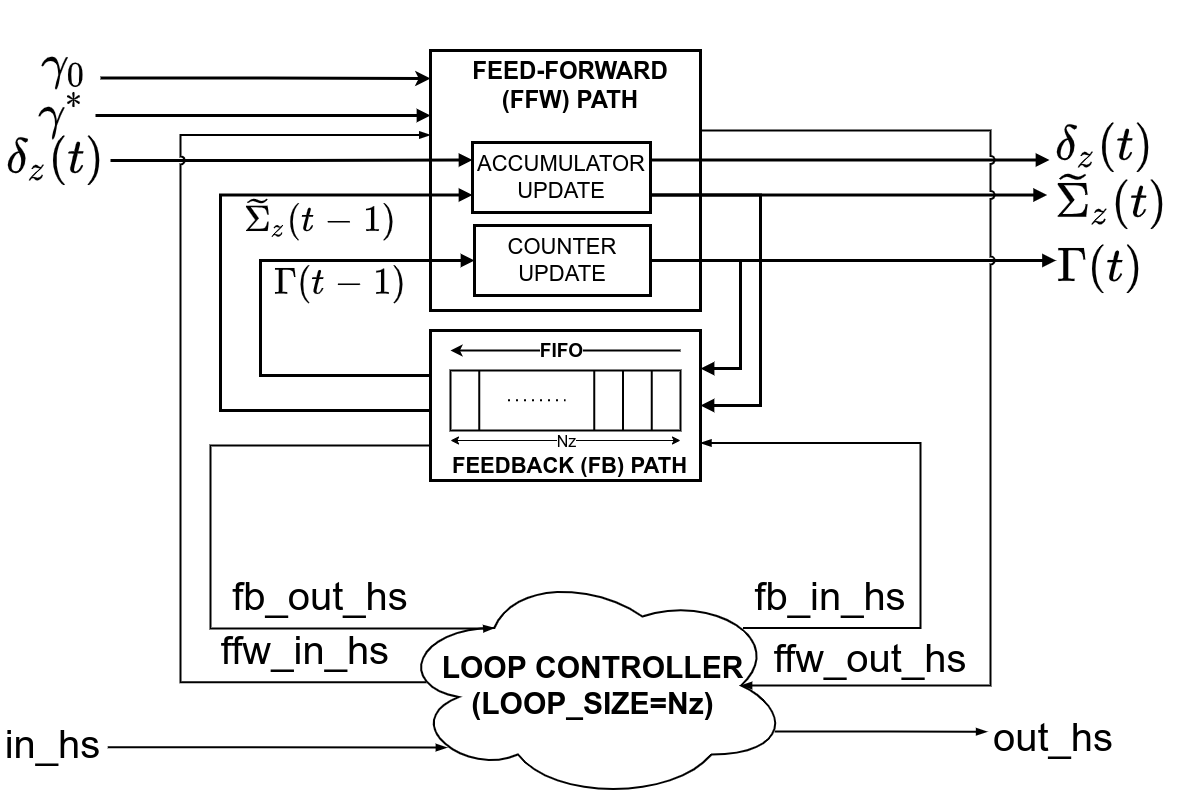}
			\caption{Code Adaptive Selection Statistics Unit top level architecture}
			\label{fig:acss_top}
		\end{figure}
		
		\begin{figure*}[ht]
			\centering
			\captionsetup{justification=centering}
			\includegraphics[width=0.95\linewidth]{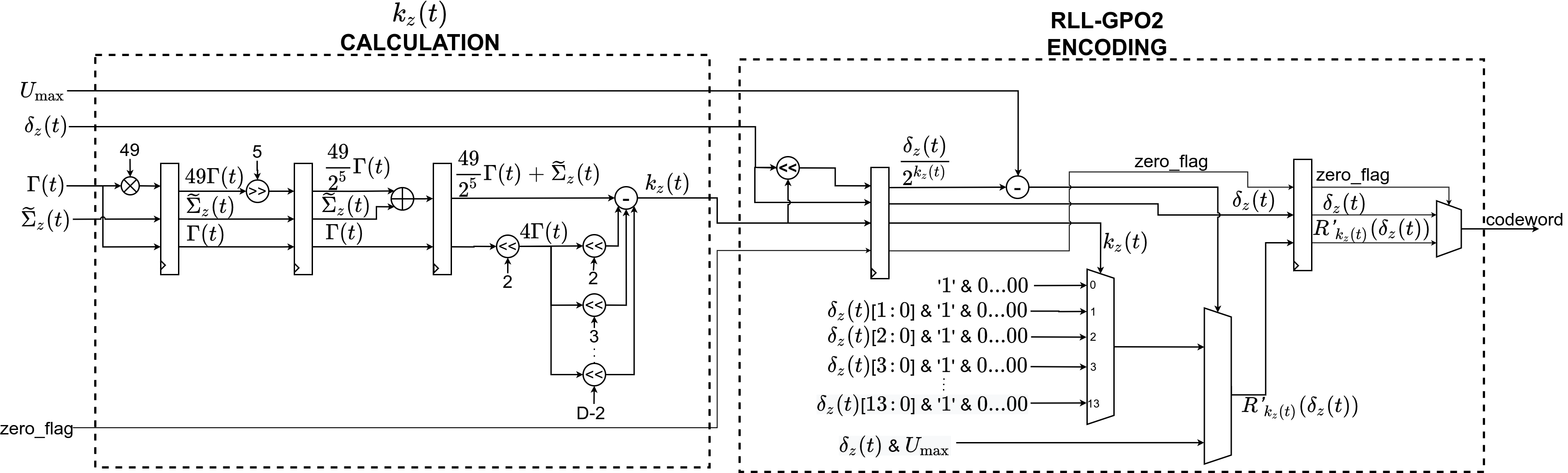}
			\caption{High Entropy Coder Unit schematic}
			\label{fig:HiEC}
		\end{figure*}
		
		\subsection{\textbf{Adaptive Code Selection Statistics Unit}}
		The \textit{Adaptive Selection Statistics} (\texttt{ACSS}) Unit maintains and updates the accumulator $\tilde{\Sigma}_z(t)$, and counter $\Gamma(t)$,
		values according to equations~(\ref{eq: sigma_rescale}) and~(\ref{eq: gamma_rescale}), supplying them to the downstream units.
		
		Both variables are updated when a new, $\delta_z(t)$, enters the encoder and rescaled when the counter reaches the value of the
		$\gamma^*$ parameter.
		Under BSQ ordering one accumulator and one counter would be required. Under BIL ordering, the same elements and resources are required for each spectral band,
		which is $N_z$ accumulators and counter values. Under BIP ordering $N_z$ accumulator values and a single counter value are required. 
		
		For our BIP implementation, the current values of accumulator and counter, $\tilde{\Sigma}_z(t)$ and $\Gamma(t)$, are computed using their previous values, $\tilde{\Sigma}_z(t-1)$ and $\Gamma(t-1)$
		respectively, creating a dependency. To resolve it, values of $\tilde{\Sigma}_z(t)$ for all bands are stored
		in a FIFO queue of depth at least equal to the number of bands, also acting as a delay buffer for the value of $\Gamma(t)$, taking advantage of the interleaved processing between
		the $N_z$ spectral bands. The feed-forward path of the loop is comprised of 2 pipeline stages, and $N_z >2$, therefore the dependency is not violated.
		
		A a similar architecture is estimated for the BIL implementation as well. In that case accumulator values would be stored in the FIFO queue at the end of every line of every band. In a BSQ architecture, the FIFO queue would not be included in the feedback datapath, acting only as a buffer for $\tilde{\Sigma}_z(t)$ values at the and of every band in order to be emptied during the construction of the compressed image-tail. Finally, \textit{Loop Controller} logic would be modified to meet each orders handshake requirements.
		
		Code Statistics calculation is architecturally very similar to the Sample Adaptive Coder of~\cite{ccsds123_b1_blue_book}, therefore an indicative implementation would be close in required resources as in~\cite{santos_shyloc_ieee_taes_2020},\cite{barrios_shyloc2_2020},\cite{shyloc_datasheet}, with the exception of the BSQ order, where the FIFO queue of $N_z$ depth would still be in use for storing $\tilde{\Sigma}_z(t)$ for the compressed image-tail. 
		
		The top-level architecture of the \texttt{ACSS} unit is shown in~\autoref{fig:acss_top}. Signals with the
		``\texttt{\_hs}'' suffix represent handshaking signals, shown in a simplified \texttt{ready/valid} notation. As a
		whole, the \texttt{ACSS} unit receives a mapped quantizer index, $\delta_z(t)$ (of $D$ bits) input and produces $\tilde{\Sigma}_z(t)$ (of $2+D+\gamma^{*}$
		bits), $\Gamma(t)$ (of $\gamma^{*}$ bits), $\delta_z(t)$ and certain flags used for codeword selection downstream. The unit consists of a feed-forward
		path which updates $\tilde{\Sigma}_z(t)$ and $\Gamma(t)$ and a feedback path, which comprises a queue storing previous
		$\tilde{\Sigma}_z(t)$ values, returning them as $\tilde{\Sigma}_z(t-1)$ to the feed-forward path.
		
		Also, for the construction of the compressed image-tail, additional logic is introduced, not shown for clarity, that activates on assertion of the end of
		image flag, in order to extract and output the $N_z$ final values of $\tilde{\Sigma}_z(t)$ from the feedback queue.
		
		\subsubsection{\textbf{Loop Controller}} \label{Loop Controler}
		The \textit{Loop Controller} is a generic IIR (Infinite Impulse Response) filter flow controller with an arbitrary
		pipelining depth in the feed-forward and feedback paths defined as RTL generics.
		There is a feed-forward pipelined path
		with $M$ pipeline registers and a feedback path with $N$ delay registers. The filter executes a function between the
		incoming samples and the feedback samples, for example
		\begin{equation}
			y(t) = \alpha x(t) + y(t-K)
			\label{eq: filter_example}
		\end{equation}
		where $K$ is the feedback dependency of the loop.
		
		The loop controller ensures that only up to $K$ samples can traverse the filter by manipulating the
		handshaking signals at the inputs and outputs of the feed-forward and feedback paths. If $K$ is less or equal than the total pipeline stages of the filter ($N+M$),
		then the loop controller inserts stall cycles in the loop, in order to not violate the data dependency, limiting data throughput to $N/(M+N)$ samples/cycle.
		Otherwise, the filter operates in a constant data rate of 1 sample/cycle.
		
		For the Hybrid Entropy Coder, the filter is described by~(\autoref{eq: sigma_rescale}). For the BIP order, the same equation can be re-written to
		resemble~(\autoref{eq: filter_example}) as
		$$	\tilde{\Sigma}(t)=\left\{
		\begin{array}{ll}
			\tilde{\Sigma}(t-N_z)+4 \delta(t) & ,\Gamma(t-N_z)<2^{\gamma^{*}}-1 \\
			{\Bigg\lfloor\dfrac{\tilde{\Sigma}(t-N_z)+4 \delta(t)+1}{2}\Bigg\rfloor} & ,\Gamma(t-N_z)=2^{\gamma^{*}}-1
		\end{array}\right.
		$$
		with $M=2$ pipeline registers in the feed-forward path and $N=N_z$ delay registers on the feedback path.
		In case of $N_z \leq 2$, stall cycles would be
		introduced by the loop controller, limiting throughput to $N_z/(N_z+2)$ samples/cycle, but this is a highly unlikely
		use-case, since there are no multispectral (or hyperspectral) images with such low number of bands (multispectral typically refers to 3 up to 15 bands). 
		
		\subsection{\textbf{High/Low Entropy Decision Unit}}
		The design of this unit revolves around the multiplication between $\Gamma(t)$ and the threshold constant $T_0$ as shown in~(\autoref{eq: hilo_sel}).
		Constant $T_0$ is not a power of 2, thus embedded multiplier blocks, should be used depending on the FPGA technology.
		When targeting Xilinx technology, a Xilinx DSP48E2 Slice is used to perform the multiplication operation, registering both inputs
		and the product with the internal DSP48E2 pipeline registers.
		After the product is calculated a comparison follows that determines the encoder choice.
		A binary \textit{high/low}(hilo) flag signals this decision, and is later used in the \textit{Codeword Combiner} unit.
		
		\subsection{\textbf{High Entropy Coder Unit}}
		In the \textit{High Entropy Coder} (\texttt{HiEC}) unit $\delta_z(t)$ is encoded with a ``high entropy'' RLL-GPO2 codeword.
		The \texttt{HiEC} unit (\autoref{fig:HiEC})
		comprises two sub-modules, the 3-stage pipelined \textit{$k_z(t)$ Calculation} module and the 2-stage pipelined
		\textit{RLL-GPO2 Encoding} module. The former
		calculates the code index $k_z(t)$ in $D-2$ bits as described in equations~(\ref{eq: kz_limit}) and~(\ref{eq: kz_high}), while the latter calculates the
		RLL-GPO2 codeword $\Re_{k_z(t)}^{\prime}\big(\delta_z(t)\big)$ in $D+U_{max}$ bits.
		
		The unit receives $\delta_z(t)$ and the code statistics $\tilde{\Sigma}_z(t)$ and $\Gamma(t)$, along with $U_{max}$ and the zero flag ($t=0$), and
		produces an RLL-GPO2 code along with its size in $8$ bits, producing 1 sample (codeword and code-size pair) per cycle.
		
		\subsection{\textbf{Low Entropy Coder Unit}}
\begin{figure}[t]
			\centering
			\captionsetup{justification=centering}
			\includegraphics[width=\linewidth]{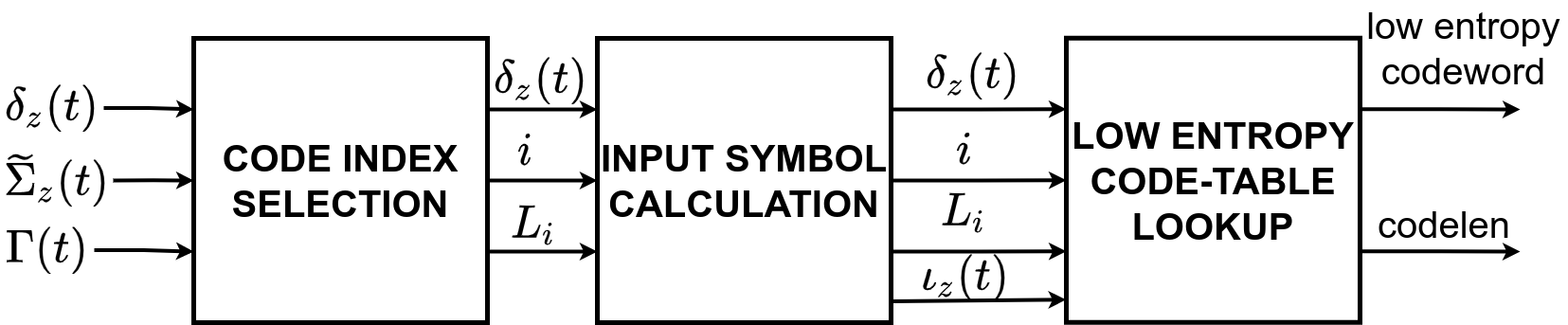}
			\caption{Low Entropy Coder Unit top level block diagram}
			\label{fig:LoEC_top}
		\end{figure}

		The \textit{Low Entropy Coder} (\texttt{LoEC}) unit is responsible for emitting output codewords from the low entropy
		code-tables, by combining multiple input symbols in a single output codeword. The
		encoding procedure is implemented by
		three sub-units as shown in~\autoref{fig:LoEC_top}.
		
		First, the code table index $i$ is selected by \textit{Code Index Selection} unit, followed by the determination of the input symbol $\iota_z(t)$ in
		\textit{Input Symbol Calculation} unit. A series of input symbols for a certain $i$, is used to search for a matching
		input codeword at the selected code-table. If one is found, the corresponding output codeword is emitted as the low
		entropy codeword along with its respective code-length. This procedure is performed by the
		\textit{Low Entropy	Code-Tables Lookup} unit. Finally, additional logic is implemented to extract flush codes from the 16
		flush code-tables during the construction of the compressed image tail, signaled by an end-of-image flag.
		
		\subsubsection{\textbf{Code Index Selection \& Input Symbol Calculation}}
		The \textit{Code Index Selection} unit selects the code-table index $i$, which will be selected for lookup. It receives
		$\delta_z(t)$ and statistics $\tilde{\Sigma}_z(t)$ and $\Gamma(t)$ as inputs and emits the code index $i$ and input symbol limit $L_i$,
		as outputs, where $i$ is the largest code index satisfying~(\autoref{eq: index_sel}).
		
		To perform the $\Gamma(t)\cdot T_i, i=0,1,...,15$ multiplications, where $T_i$ are not powers-of-two, embedded multiplier blocks are used
		(e.g. 3-stage pipelined DSP48E2 Slices in Xilinx technology). All 16 multiplications are performed in parallel, and a comparison scheme selects the code index and input symbol limit.
		
		The \textit{Input Symbol Calculation} unit determines the input symbol, $\iota_z(t)$ to be used for the code-table
		lookup as described in~(\autoref{eq: sym_calc}).
		
		\begin{figure*}[ht]
			\centering
			\captionsetup{justification=centering}
			\includegraphics[width=\linewidth]{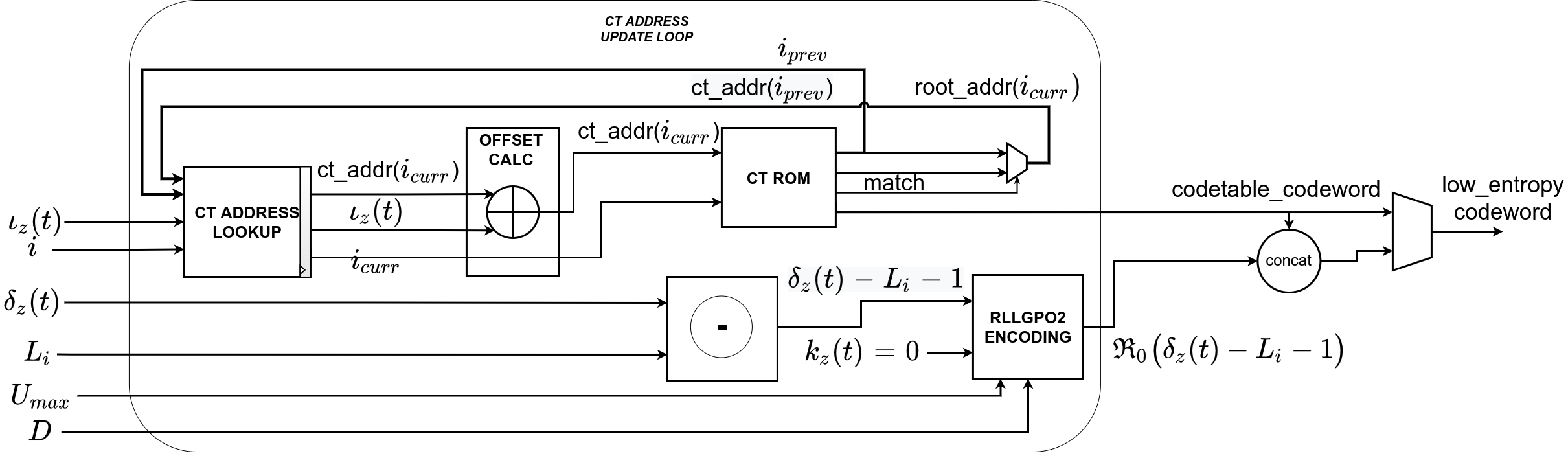}
			\caption{Low Entropy Code-Table Lookup implementation}
			\label{fig:LoEC_loop}
		\end{figure*}
		
		\subsubsection{\textbf{Low Entropy Code-Table Lookup}}
		The \textit{Low Entropy Code-Table Lookup} unit is the most complex unit and the primary
		performance bottleneck of the design. The unit contains two parallel data paths. The first path provides the low entropy codeword as a lookup to the selected code-table,
		while the second path calculates the RLL-GPO2 codeword,
		$\Re_{0}^{\prime}\big(\delta_z(t) - L_i - 1\big)$, when the input symbol is the escape symbol. The two
		codewords are concatenated, and a multiplexer selects either the single low entropy codeword, or the concatenated
		RLL-GPO2 codeword with the low entropy codeword.
		
		\paragraph{\textbf{Low Entropy Code-Tables ROM}}
		$\newline$
		A major design consideration is implementing the 16 low entropy code-tables and flush-tables efficiently.
		Taking advantage of the code-table tree structure, we adopt the representation of them in Code-Table ROMs for efficient
		lookups introduced for the first time in \cite{chatziant_obpdc2020}. The structure used to represent the code and
		flush-tables resembles a prefix-free Trie data structure.
		The tree root is the ``null'' sequence while every child-node is connected to its parent with an edge representing an input symbol.
		Terminal nodes are output codewords and non-terminal nodes are flush codewords. A sequence of input symbols that leads
		to a terminal leaf node during encoding, causes a match codeword to be output.
		To map this sequence to ROM addresses, a model of the Code-Table ROMs was developed in software to appropriately order the code-table contents to allow lookup via address increments.
		
		\begin{table}
			\centering
			\captionsetup{justification=centering}
			\caption{Example of low entropy code-table and flush table along with their tree and ROM representations }
			$\begin{array}{p{0.45\linewidth}|p{0.45\linewidth}}
				\toprule
				\multicolumn{2}{c}{\textbf{\texttt{Example code-table}}} \\
				\toprule
				\textbf{\texttt{Input codeword}} & \textbf{\texttt{Output codeword}} \\
				\midrule
				\texttt{0}  & \texttt{4'hA} \\
				\texttt{X}  & \texttt{5'hB} \\
				\texttt{10} & \texttt{4'hC} \\
				\texttt{11} & \texttt{8'hD} \\
				\texttt{1X} & \texttt{6'hE} \\
				\midrule
				\multicolumn{2}{c}{\textbf{\texttt{Example flush table}}} \\
				\midrule
				\textbf{\texttt{Active prefix}} & \textbf{\texttt{Flush word}} \\
				\midrule
				\texttt{(null)} & \texttt{1'h0} \\
				\texttt{1} & \texttt{2'h1} \\
				\bottomrule
				\toprule
				\multicolumn{2}{c}{\textbf{\texttt{Example code-table tree}}} \\
				\toprule
				\multicolumn{2}{c}{
					\includegraphics[width=0.5\linewidth]{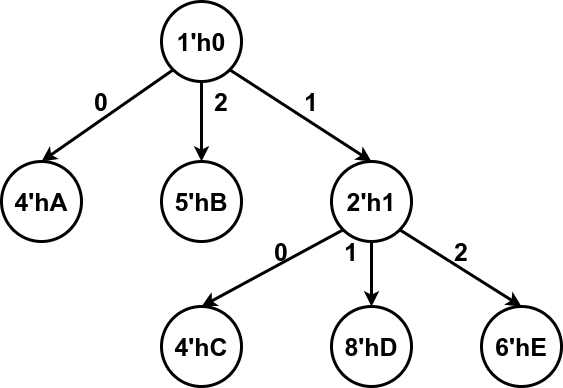}
				}\\
				\bottomrule
				\toprule
				\multicolumn{2}{c}{\textbf{\texttt{Example code-table ROM}}}\\
				\midrule
				\textbf{\texttt{Address}} & \textbf{\texttt{Data}} \\
				\midrule
				\texttt{0} & (\texttt{1'h0}, \texttt{4'hA})  \\
				\texttt{1} & (\texttt{1'h0}, \texttt{ptr = 3}) \\
				\texttt{2} & (\texttt{1'h0}, \texttt{5'hB}) \\
				\texttt{3} & (\texttt{2'h1}, \texttt{4'hC})  \\
				\texttt{4} & (\texttt{2'h1}, \texttt{8'hD}) \\
				\texttt{5} & (\texttt{2'h1}, \texttt{6'hE}) \\
				\bottomrule
			\end{array}$
			\label{tab:example_ct}
		\end{table}
		
		If input samples are exhausted before the tree for a code-table is fully traversed, then a flush codeword is emitted. Such
		codewords correspond to each non-terminal tree node and form part of the compressed image tail, which is emitted when sample encoding has finished. The Code-Table
		ROMs are looked-up with the last stored address of each tree in sequence of increasing code index, and the corresponding
		flush codeword is emitted, either from an intermediate node or from the tree's root address.
		
		To demonstrate, a similar but smaller code and flush-table are, along with the corresponding tree shown in~\autoref{tab:example_ct}. This code-table is
		transformed for efficient lookups into the code-table ROM shown at the lower part of the table, each cell containing a tuple that corresponds to a tree node.
		The first tuple element represents the node's parent flush codeword. In terminal nodes the second tuple element contains its output codeword.
		Otherwise, non-terminal nodes contain an offset which lead to the next node in the tree walk, when added to the
		incoming input symbol. The ROM contents are produced by representing each code-table with a Trie in software and then
		traversing it breadth-first, to produce appropriate pointer offsets for all possible walks from the root to the
		terminal nodes.
		
		To implement the code-table ROM scheme, we store 16 ROM address pointers, in the \textit{CT\_ADDRESS\_LOOKUP} memory shown in~\autoref{fig:LoEC_loop}. Incoming symbols are added to the previous pointer for their corresponding
		code index to form a ROM address. When an escape symbol appears, a codeword emission is guaranteed from the selected code-table and the ROM address
		is reset to the root address.
		After the image is fully encoded, additional logic handles flush codewords for the image tail construction.
		
		\paragraph{\textbf{Low Entropy Feedback Loop}}
		$\newline$
		Low Entropy feedback loop is implemented as shown in~\autoref{fig:LoEC_loop}. The selected code-table address is read from the \textit{CT\_ADDRESS\_LOOKUP}
		registers and updated by adding the current input symbol. Then it is used for reading the code-table ROM (\texttt{CT\_ROM}) and finally is written
		back to the registers.
		
		In addition to the low entropy codeword, whenever the input symbol is the escape symbol, an additional RLL-GPO2 codeword for $\delta_z(t)-L_i-1$
		is produced by this unit.
		Therefore there are two output codewords corresponding to a single input.
		
		The code-table address update procedure is performed in a single clock cycle providing full-throughput of one sample/cycle. The critical path of the
		design is also located in the feedback loop and defines the maximum achievable frequency.
		
		\subsection{\textbf{Codeword Combiner Unit \& Variable Length Code Packer Unit}}
		$\newline$
		Depending on the \textit{HiLo} decision flag and certain other flags, \textit{Code Combiner} unit emits the appropriate codeword into
		the Variable Length Code (VLC) Packer's input, which produces fixed 64-bit packets.
		In the case of a low entropy codeword, if the input symbol is the escape symbol, then the high
		entropy codeword must precede the low entropy code-table output codeword, meaning that there are two codewords to
		forward to the VLC packer corresponding to the same input sample. Also, whenever $\Gamma(t)$ rescales, the least
		significant bit of $\tilde{\Sigma}_z(t)$ is emitted to the bitstream for decoding purposes, meaning that it should precede the current
		codeword output.
		
		The \textit{Codeword Combiner} unit handles those special cases using flags produced throughout the encoding to provide the required output codewords
		to the VLC packer, as well as handling the sequence of outputs that constitute the compressed image tail.
		
		After the \textit{Codeword Combiner} unit has extracted the appropriate codeword, the
		\textit{Variable Length Code (VLC) Packer} unit accepts variable-length codewords as inputs along with their length, and
		packs them into a 64-bit packets comprising the final bitstream.
		This component is re-used from previous work in~\cite{tsigkanos_123b1_2018} and is capable of operating in high data rates.
		
		\begin{table}[t]
			\centering
			\caption{Implementation statistics for XCKU040 FPGA}
			$\begin{array}[width=\linewidth]{l@{}cc}
				\toprule
				\textbf{Image} & \begin{tabular}{@{}c@{}}AVIRIS\\(680$\times$512$\times$224,16bpppb)\end{tabular} &
				\begin{tabular}{@{}c@{}}AVIRIS-NG\\(640$\times$512$\times$432,14bpppb)\end{tabular} \\
				\midrule
				\textbf{Frequency} & \multicolumn{2}{c}{305\mbox{ MHz}}\\
				\midrule
				\textbf{Est. Power} & \multicolumn{2}{c}{1.525\mbox{ W}}\\
				\midrule
				\begin{array}{@{}l@{}}\textbf{Device} \\ \textbf{utilization}\end{array} &
				\begin{array}{@{}c@{}}
					5086 (2.05\%) \mbox{ LUTs} \\
					1 (0.08\%) \mbox{ BRAMs} \\
					17 (0.89\%) \mbox{ DSPs} \\
					3296 (0.67\%) \mbox{ FFs}
				\end{array} &
				\begin{array}{@{}c@{}}
					5067 (2.09\%) \mbox{ LUTs} \\
					1 (0.08\%) \mbox{ BRAMs} \\
					17 (0.89\%) \mbox{ DSPs} \\
					3301 (0.68\%) \mbox{ FFs}
				\end{array} \\
				\midrule
				\textbf{MSamples/sec} & \multicolumn{2}{c}{305}  \\
				\midrule
				\textbf{Gbps (@16bpppp)} & \multicolumn{2}{c}{4.88} \\
				\bottomrule
			\end{array}$
			\label{tab:impl_res2}
		\end{table}

		\section{\textbf{Experimental Results}}
		The proposed CCSDS-123.0-B-2 Hybrid Entropy Coder architecture implemented as an IP core was verified using
		simulation-based (RTL) and FPGA-in-the-loop (FIL) based verification to speed-up verification process on a ZedBoard FPGA
		development board against a software golden model in Python, developed and provided by Universitat Autònoma de Barcelona (UAB).
		
		The proposed architecture is implemented with the encoder parameters (\mbox{\autoref{tab:params}})
		defining the image dimensions ($N_x$, $N_y$, $N_z$), sample dynamic range ($D$), Unary
		Length limit ($U_{max}$), Initial Count Exponent ($\gamma_0$) and the Rescaling Counter Size ($\gamma^{*}$).
		The encoder can be configured with run-time configurable parameters through a memory-mapped register interface, while VHDL generics are used to constrain the parameters' maximum allowable range.
		Using $N_x$ as an example, at netlist generation time
		(compile-time) a VHDL generic \texttt{g\_Nx\_max} sets the maximum usable number of image columns and then at run-time through
		the configuration interface, this instance of the IP Core can be configured for values of $N_x$ to compress images with
		$N_x < \texttt{g\_Nx\_max}$. This feature allows tailoring to optimize the design by minimizing resource utilization or increasing
		achievable frequency, at the expense of increased complexity in the RTL architecture and design. In all cases, the generics should not get values that exceed the maximum allowed value of the corresponding parameter as defined by the Standard~\cite{ccsds123_b2_blue_book}.
		
		\begin{table}[t]
			\centering
			\caption{Hybrid Entropy Coder list of parameters}
			$\begin{array}{p{0.2\linewidth}|p{0.2\linewidth}|p{0.45\linewidth}}
				\toprule
				\mbox{Parameter} & \mbox{Support} & \mbox{Range} \\
				\midrule
				$N_x$ & \mbox{yes} & 2\mbox{ up to g\_Nx\_max} \\
				$N_y$ & \mbox{yes} & 2\mbox{ up to g\_Ny\_max} \\
				$N_z$ & \mbox{yes} & 3\mbox{ up to g\_Nz\_max} \\
				$D$   & \mbox{yes} & 4\mbox{ up to g\_D\_max} \\
				$\gamma_0$   & \mbox{yes} & 1\mbox{ up to g\_go\_max} \\
				$\gamma^{*}$ & \mbox{yes} & $\max\{4, \gamma_0+1 \}$\mbox{ up to g\_gs\_max} \\
				$U_{max}$    & \mbox{yes} & 8\mbox{ up to g\_Umax\_max} \\
				\bottomrule
			\end{array}$
			\label{tab:params}
		\end{table}
		
		\subsection{\textbf{Design Verification}}
		The Hybrid Entropy Coder design was verified using simulation-based verification at VHDL RTL with Mentor Questa against the	software golden model to ensure functional coverage of all corner cases and also targeting high VHDL code coverage (statement, branch, FSM and condition). The testing framework is based on the VUnit~\cite{VUnit} Python testing framework
		and a set of python scripts. Test campaigns comprising of different images,	compile time (generics) and run-time (compression) parameters are described in test files in this framework. The test scripts interpret the parameters to
		invoke the golden compressor binary to produce the verification data. Then, a testbench implemented as pass/fail test
		instrumented with VUnit is invoked with the Questa simulator with automatic checking comparing with golden responses.
		
		A comprehensive test suite exercising all combinations of the encoder's parameters
		($Umax$,$\gamma_0$,$\gamma*$) was used to verify functional correctness.
		Finally, tests incorporating full images from AVIRIS~\cite{aviris} and AVIRIS-NG~\cite{aviris-ng} image sets and synthetic
		images to debug corner case scenarios were applied to verify the encoder against realistic use-case scenarios.
		
		More comprehensive verification was performed using FPGA-in-the-loop (FIL) techniques on a ZedBoard FPGA development board
		hosting a Xilinx Zynq-7000 SoC FPGA device, also leveraging the ARM embedded processor. For the purposes of FIL
		verification, several hyperspectral test cubes including synthetic and random test images and using multiple configurations
		were transferred to the board and the compressed images was received from the board using a JTAG-to-AXI interface.
		
		\subsection{\textbf{Design Implementation}}
		The CCSDS-123.0-B-2 Hybrid Entropy Coder was implemented targeting Xilinx Kintex-Ultrascale technology (XCKU040-2FFVA1156E
		FPGA) and is therefore directly transferable (in terms of implementation results), to the Xilinx Radiation Tolerant XQRKU060
		FPGA. Implementation on the target device was performed using a configuration for the AVIRIS (680$\times$512$\times$224,
		16bpppb), AVIRIS-NG (640$\times$512$\times$432, 14bpppb) hyperspectral instruments, which are typical hyperspectral
		sensors and the standard benchmark in the literature.
		
		\autoref{tab:impl_res2} presents implementation (Place \& Route, Timing Analysis) results for the Kintex Ultrascale FPGA
		using Xilinx Vivado 2020.2. The implementation parameters used are those suggested as defaults
		in~\cite{blanes_parameters_tuning_2019} and~\cite{ccsds123_green_book} ($U_{max}=18, \gamma_0=1$, $\gamma^*=6$).
		For more accurate implementation results, the generic parameters defining the maximum allowed values of encoder inputs,
		are set to be equal to the exact input parameter value.
		
		The proposed architecture achieves a constant throughput of $\sim$305 MSamples/sec operating at 305 MHz while the FPGA resource footprint is kept low.
		The power consumption is reported for the whole FPGA including SpaceFibre interface IP cores and the Kintex Ultrascale device GTH transceivers.
		
		\begin{figure}[t]
			\centering
			\includegraphics[width=0.95\linewidth]{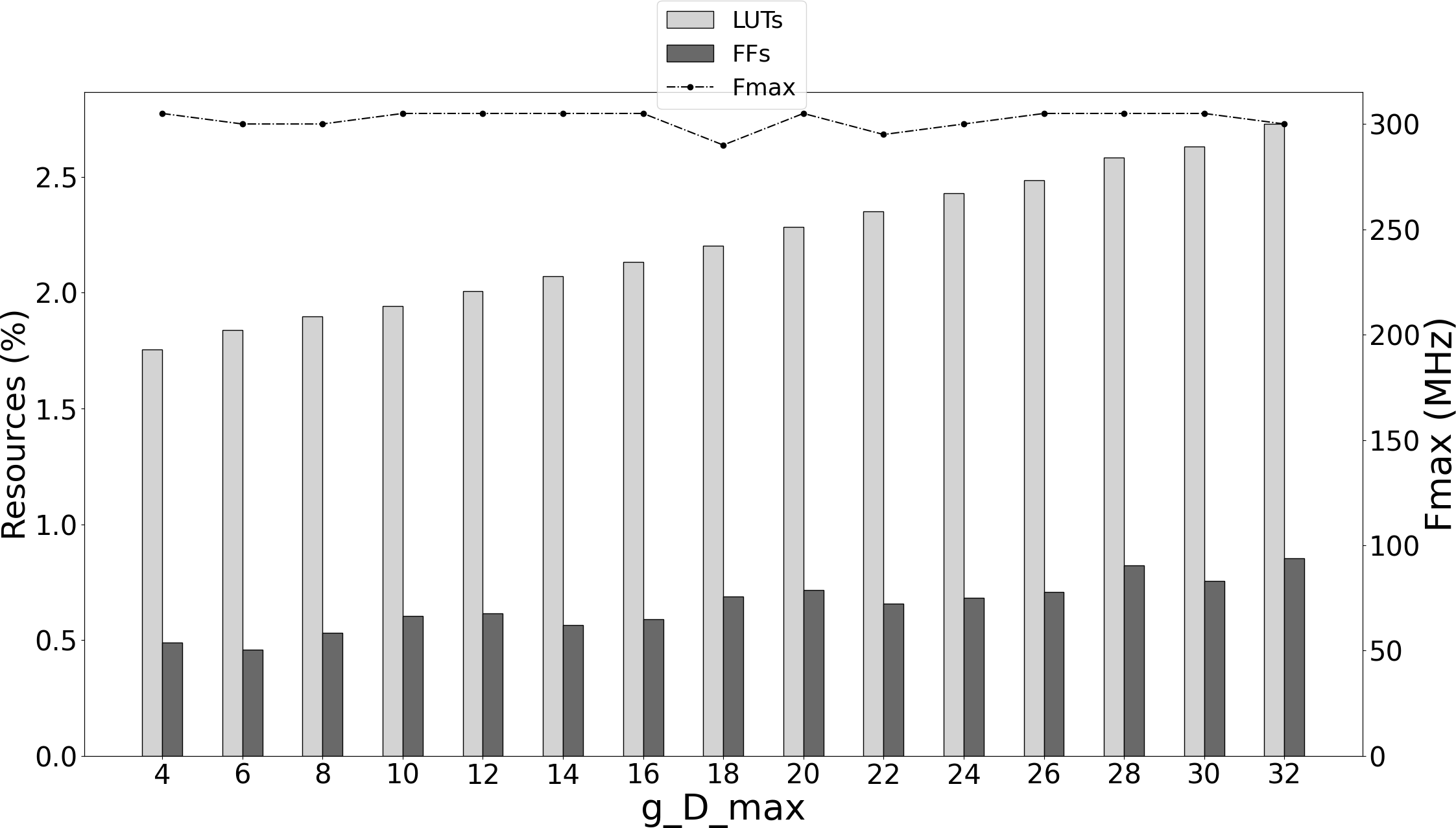}
			\caption{Maximum frequency and resource usage with respect to \texttt{g\_D\_max}}
			\label{fig:single_D_Fmax}
		\end{figure}
		
		Design's data rate throughput can be estimated by the ratio between total image samples and total clock cycles,
		needed for a complete encoding.
		\begin{small}
			\begin{equation}
				\begin{array}{l}
					data\_rate = \dfrac{N_xN_yN_z}{init + N_xN_yN_z + 16 + N_z + T_{esc\_syms}}
				\end{array}
				\label{eq: data_rate_single}
			\end{equation}
		\end{small}
		The total number of samples are divided by the total number of clock cycles. Here,
		$init$, are the initial cycles required for the pipeline to fill up and for the header generation.
		In addition to these initial cycles, there are $N_x \cdot N_y \cdot N_z$ samples processed in one cycle each and the cycles consumed
		for the image tail creation, which are the extraction of 16 flush-codewords followed by $N_z$ accumulator values also
		processed in one cycle each.
		Finally, $T_{esc\_syms}$, is the number of escape symbols that force the low entropy coder to produce an additional RLL-GPO2
		codeword to the low entropy code-table codeword, requiring an additional clock cycle in encoding.
		
		\autoref{fig:single_D_Fmax} presents the resource usage and maximum frequency after Place \& Route for the
		generic-configurable parameter \texttt{g\_D\_max}.
		Maximum frequency of the IP Core is slightly influenced by the increase of \texttt{g\_D\_max} estimated between 290 MHz and 305 MHz for \texttt{g\_D\_max} = 2,4,6,...,32, while resources tend to increase as data-path width is close relating to \texttt{g\_D\_max}.
		
		The throughput performance is stable, providing a constant data rate of $\sim$1 sample/cycle
		which does not depend on the hyperspectral image data statistics and is not degraded when high
		Absolute Error Limit Constants are configured leading to a large number of low entropy encoded
		samples as in our preliminary work\cite{chatziant_obpdc2020}. The critical path of the design is located in the low entropy coder's
		feedback loop datapath which determines the maximum operational frequency. The code-table ROM component included in this
		feedback loop path is implemented using asynchronous distributed RAM (LUTRAM), instead of BRAM in order to avoid the RAW hazard of \texttt{LoEC} loop, resulting in increased LUT usage.
		
		\begin{table}
			\centering
			\caption{Data-rate of the Hybrid Entropy Coder IP Core validated for different hyperspectral images}
			$\begin{array}[width=0.95\linewidth]{l@{}ccccc}
				\toprule
				\multirow{2}{*}{\mbox{\textbf{Image}}} & \multirow{2}{*}{\mbox{\textbf{A*}}} & \multirow{2}{*}{\textbf{N$_x$}} & \multirow{2}{*}{\textbf{N$_y$}} & \multirow{2}{*}{\textbf{N$_z$}} & \mbox{\textbf{Throughput}}\\
				& & & & & \mbox{\textbf{(samples/cycle)}}\\
				\midrule
				\mbox{AVIRIS\_yellowstone\_sc00} & $0$ & 512 & 680 & 224 & 0.999\\
				\mbox{AVIRIS\_hawaii} & $0$ & $512$ & $614$ & $224$ & 0.989\\
				\mbox{AVIRIS\_maine}  & $0$ & $512$ & $680$ & $224$ & 0.988\\
				\mbox{AVIRIS-NG\_A} & $0$ & $512$ & $640$ & $432$ & 0.999\\
				\mbox{AVIRIS-NG\_A} & $1$ & $512$ & $640$ & $432$ & 0.990\\
				\mbox{AVIRIS-NG\_B} & $0$ & $512$ & $640$ & $432$ & 0.999\\
				\mbox{AVIRIS-NG\_B} & $1$ & $512$ & $640$ & $432$ & 0.990\\
				\mbox{PRISMA\_land} & $0$ & $1000$ & $1000$ & $173$ & 0.999\\
				\mbox{PRISMA\_land} & $1$ & $1000$ & $1000$ & $173$ & 0.987\\
				\mbox{PRISMA\_land} & $2$ & $1000$ & $1000$ & $173$ & 0.991\\
				\mbox{PRISMA\_ice} & $0$ & $1000$ & $1000$ & $173$ & 0.999\\
				\mbox{PRISMA\_ice} & $1$ & $1000$ & $1000$ & $173$ & 0.990\\
				\mbox{PRISMA\_ice} & $2$ & $1000$ & $1000$ & $173$ & 0.990\\
				\mbox{PRISMA\_ocean} & $0$ & $1000$ & $1000$ & $173$ & 0.999\\
				\bottomrule
			\end{array}$
			\label{tab: validation_res}
		\end{table}
		
		\subsection{\textbf{Design Validation}}
		The CCSDS-123.0-B-2 Hybrid Entropy Coder's validation and demonstration set-up is built around SpaceFibre (ECSS-E-ST-50-11C) test equipment, provided by STAR-Dundee, to interface the Xilinx KCU105 development board and match standard space deployment.
		SpaceFibre is a very high-speed (5 Gbit/s) serial link and network technology, designed specifically for use on board spacecraft. 
		
		The CCSDS-123.0-B-2 Hybrid Entropy Coder validation and demonstrator set-up includes a standard PC emulating Electronic Ground Support
		Equipment (EGSE). The EGSE PC hosts a STAR-Ultra PCIe board which is connected to the KCU105 development board using QSFP to SFP+ cable assembly. SpaceFibre interface
		VHDL IP Cores are also implemented in the XCKU040 FPGA hosted in KCU105 board to provide AXI4-Stream interface with the CCSDS-123.0-B-2 Hybrid Entropy Coder IP Core data
		inputs and compressed output over a singe data Virtual Channel (VC). A single lane SpaceFibre link able to provide 6.25 Gbps (effective 5.0 Gbps) data-rate is sufficient
		for the validation and demonstration of the CCSDS-123.0-B-2 Hybrid Entropy Coder.
		
		A large set of test images from the CCSDS corpus of images~\cite{ccsds_image_corpus} along with several images from the
		PRISMA Hyperspectral mission~\cite{prisma} launched March 2019, with multiple compression configurations was applied by the EGSE PC
		for the validation of the Hybrid Entropy
		Coder. \autoref{tab: validation_res}
		displays the achieved throughput for some notable images used in the validation process. These figures were validated by SpaceFibre link analyser
		software installed in the EGSE PC along with performance counters instrumenting the IP Core in the XCKU040 FPGA.
		Images were compressed with in lossless ($A*=0$) and near-lossless ($A*>0$) mode, always operating at $\sim$1 sample-per-cycle efficiency in agreement with the data rate estimation of~\autoref{eq: data_rate_single}.
		The presented CCSDS-123.0-B-2 Hybrid Entropy Coder IP Core achieved 305 MSamples/sec (4.88 Gbps) throughput performance.
		
		\begin{table*}[ht]
			\centering
			\caption{Comparison with implementation of FLEX's Hybrid Entropy Coder}
			$\begin{array}[width=\linewidth]{lcccccc}
				\toprule
				\multirow{2}{*}{\mbox{\textbf{Implementation}}} & \multirow{2}{*}{\mbox{\textbf{LUTs}}} & \multirow{2}{*}{\mbox{\textbf{FFs}}} & \multirow{2}{*}{\mbox{\textbf{DSP48Es}}} & \multirow{2}{*}{\mbox{\textbf{BRAMs}}} & \mbox{\textbf{Frequency}} & \mbox{\textbf{Throughput}}\\
				& & & & & \mbox{\textbf{(MHz)}} & \mbox{\textbf{(Cycles/Sample)}}\\
				\midrule
				\mbox{FLEX Hybrid Entropy Coder} & \multirow{2}{*}{3341} & \multirow{2}{*}{1293} & \multirow{2}{*}{16} & \multirow{2}{*}{27} & \multirow{2}{*}{168.8} & \multirow{2}{*}{7} \\
				\mbox{\cite{keymeulen_flex_2016},\cite{keymeulen_ngis_2018},\cite{keymeulen_ngis_presentation_2018}}\\
				\midrule
				\mbox{This work} & 5085 & 3322 & 17 & 1 & 305 &  1\\
				\bottomrule
			\end{array}$
			\label{comparison}
		\end{table*}
		
		\section{\textbf{Comparison With Previous Work}}
		The previous issue of the standard, CCSDS-123.0-B-1, describes only lossless compression and is considered a mature solution for on-board hyperspectral compression.
		Issue 2, shares many implementation similarities to Issue 1, regarding the lossless compression option, therefore implementation of Issue 1 are considered comparable prior work.
		Multiple implementations have been presented in the literature designed for various trade-offs and  devices such as FPGAs and GPUs, as well as Systems-on-a-chip (SoC). SHyLoC 1.0 \cite{santos_shyloc_ieee_taes_2020} and SHyLoC 2.0~\cite{barrios_shyloc2_2020} implementations, available at
		the European Space Agency (ESA) IP Cores library to be licensed for space missions, research and/or commercial use, under specific conditions, provide a feature
		complete implementation of CCSDS-123.0-B-1 and CCSDS-121.0-B-2 algorithms, as a technology agnostic IP Core suitable for FPGA and ASIC technologies. Moreover, SHyLoC IP
		Cores provide wide and versatile parameterization and configuration options enabling reduced complexity and footprint when dealing with FPGA devices with a limited
		amount of resources.
		Other implementations provide high-throughput by using a either a single compression
		engine,~\cite{tsigkanos_123b1_2018},~\cite{tsigkanos_123b1_obpdc2018},~\cite{bascones_fpga_2018},~\cite{fjeldtvedt_123b1_2018} and leveraging the interleaved processing
		of BIP pixel order format that enables deep pipelining presented for the first time in~\cite{theodorou_obpdc2016} or by exploiting the CCSDS-123.0 image segmentation
		and task-level parallelism along with Commercial Off-the Shelf (COTS) FPGA SoC
		technologies~\cite{bascones_parallel_2017},~\cite{rodriguez_SoC_2019},\cite{orlandic_123b1_parallel_2019}, achieving state-of-the-art throughput performance
		~\cite{tsigkanos_SoC_2020}. Implementation on GPU devices ~\cite{ferraz_15gbits_2020},\cite{ferraz_gbits_2021},\cite{ferraz_hyperspectral_2021} utilize GPUs and
		heterogeneous CPU and GPU systems to parallelize the CCSDS-123.0-B-1 standard by exploiting image segmentation and task-level parallelism, achieving very high-throughput,
		but higher energy consumption when compared to FPGA implementations.
		
		Due to the recent release of Issue 2, there are few known implementations of CCSDS-123.0-B-2 in literature to date, none of which involved the VHDL RTL
		implementation of the Hybrid Entropy Coder or the full CCSDS-123.0-B-2 standard.
		
		In \cite{lopez_cueva_evaluation_2021} the authors present a parallel implementation in software of the near-lossless CCSDS-123.0-B-2 standard for the evaluation of the
		RC64 many-core rad-hard processor~\cite{ginosar_rc64_2016}. However, they implement only the Sample Adaptive Entropy Coder while the Hybrid Entropy Coder was not considered due to implementation challenges related to throughput performance. This parallelization scheme achieves high speed-up when all 64 cores are used, with maximum throughput of 0.45 MSamples/sec, and limited performance when there are idle cores.
		
		In~\cite{trohin_parallelization_2020}, the authors propose parallel implementations of both Issue 1 and Issue 2 of CCSDS-123.0-B-2 in software with OpenMP targeting
		different space qualified CPUs (i.e. GR740, LS1046). Their work suggests ways of splitting data and assigning jobs among the available CPU cores, for both lossless and 	near-lossless predictor and hybrid entropy encoder.
		
		The Fast Lossless Extended (FLEX) algorithm ~\cite{keymeulen_flex_2016},\cite{keymeulen_ngis_2018},\cite{keymeulen_ngis_presentation_2018} is the algorithmic basis for CCSDS-123.0-B-2 and the
		Hybrid Entropy Coder is an extension of the FLEX's original hybrid entropy coder, therefore FLEX implementations can be considered for comparison purposes. Experimental
		results for the FLEX entropy coder targeting the Virtex 5 FX130T FPGA technology reach a maximum frequency of 168.8 MHz and a throughput of 24 MSamples/sec using (7
		cycles/sample). For the whole FLEX compressor a throughput of 3.4 MSamples/sec was achieved at 82.5 MHz maximum frequency (24 cycles/sample). Although a direct comparison with FLEX entropy coder in terms of maximum frequency is not appropriate because this paper considers a next-generation space-grade FPGA platform, the proposed Hybrid Entropy Coder architecture achieves 7 times higher throughput performance in terms of samples/cycle.
		~\autoref{comparison} summarizes the comparison of RTL implementation of FLEX Hybrid Entropy Coder 
		with the presented work.
		
		The first, full implementation of the CCSDS-123.0-B-2 standard, although using High-Level Synthesis (HLS), was presented in~\cite{barrios_implementation_2020}. The
		CCSDS-123.0-B-2 compressor developed for the ESA CHIME space mission includes a High-Level Synthesis (HLS) implementation of the near-lossless predictor and re-uses the
		VHDL RTL implementation of the Block-Adaptive encoder as implemented for the SHyLoC~\cite{santos_shyloc_ieee_taes_2020} IP Core. The Hybrid Entropy Encoder was not
		considered, and thus, comparisons are not appropriate. The compressor in~\cite{barrios_implementation_2020} meets CHIME mission requirements of data rate
		up to 2 Gbps (@16bppb, 125 MHz), with the HLS-generated near-lossless Predictor requiring more than 1 cycles/sample, which the authors plan to improve in future implementations
		in VHDL RTL.

		\section*{\textbf{Conclusion}}
		In this paper, we introduced an efficient architecture and a high-throughput hardware implementation of the CCSDS-123.0-B-2 Hybrid Entropy
		Coder. The introduced architecture exploits the systolic design pattern to provide modularity and latency insensitivity in a deep and
		elastic pipeline, as well as an innovative approach on the Low Entropy Coder's codetable lookup design, and achieves a constant high-throughput implementation in space-grade SRAM FPGA technology (~305 MSamples/s operating at 1
		sample/cycle) with a small FPGA resource footprint. 
		The introduced architecture is validated and demonstrated on a commercially available
		Xilinx KCU105 development board hosting a Xilinx Kintex Ultrascale XCKU040 SRAM FPGA, and thus, is directly transferable to the Xilinx
		Radiation Tolerant Kintex UltraScale XQRKU060 space-grade devices for space deployments. Moreover, state-of-the-art SpaceFibre
		(ECSS-E-ST-50-11C) interface and test equipment were used in the validation platform to match space deployment. To the best of our
		knowledge, this is the 	first published fully-compliant architecture and high-throughput implementation of the CCSDS-123.0-B-2 Hybrid
		Entropy Coder, also targeting space-grade FPGA technology.
		
		\section*{Acknowledgment}
		We would like to thank Ian Blanes and Joan Serra-Sagristà from Universitat Autònoma de Barcelona (UAB) for providing the software golden
		model for the CCSDS-123.0-B-2 algorithm.
		Part of this research has received funding from the Hellenic Foundation for Research and Innovation (HFRI) and the General 
		Secretariat for Research and Technology (GSRT) under the 1st call for H.F.R.I. Research Projects for the support of
		Post-doctoral Researchers under grant agreement No 990 and part of it has received funding from the European Union’s
		Horizon 2020 research and innovation programme under grant agreement No 776151.
		
		\ifCLASSOPTIONcaptionsoff
		\newpage
		\fi

		
		
		\bibliographystyle{IEEEtran}
		\bibliography{TAES_references}
		%
		
		%
		\vskip 0pt plus -1.26fil	
		
		 \begin{IEEEbiography}[{\includegraphics[width=1in,height=1.25in,clip,keepaspectratio]{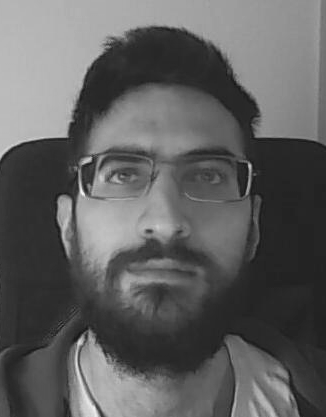}}]{Panagiotis Chatziantoniou}{\space}received the B.Sc in Computer Engineering from the Computer Engineering and Informatics Department (CEID) of University of Patras (UoP), in 2006 and M.Sc in Computer Systems: Software and Hardware from the the Department of Informatics \& Telecommunications of the National and Kapodistrian University of Athens (NKUA) at 2020. \par 
		 Currently, he is working towards his PhD degree at NKUA, Greece. He is a member of Digital Systems \& Computer Architecture Laboratory (DSCAL) of NKUA. His research interests include hardware design on-board payload data processing systems, FPGA-based acceleration, dependable and reconfigurable computing and reliability.
		\end{IEEEbiography}
		
		\vskip 0pt plus -1.26fil

		\begin{IEEEbiography}[{\includegraphics[width=1in,height=1.25in,clip,keepaspectratio]{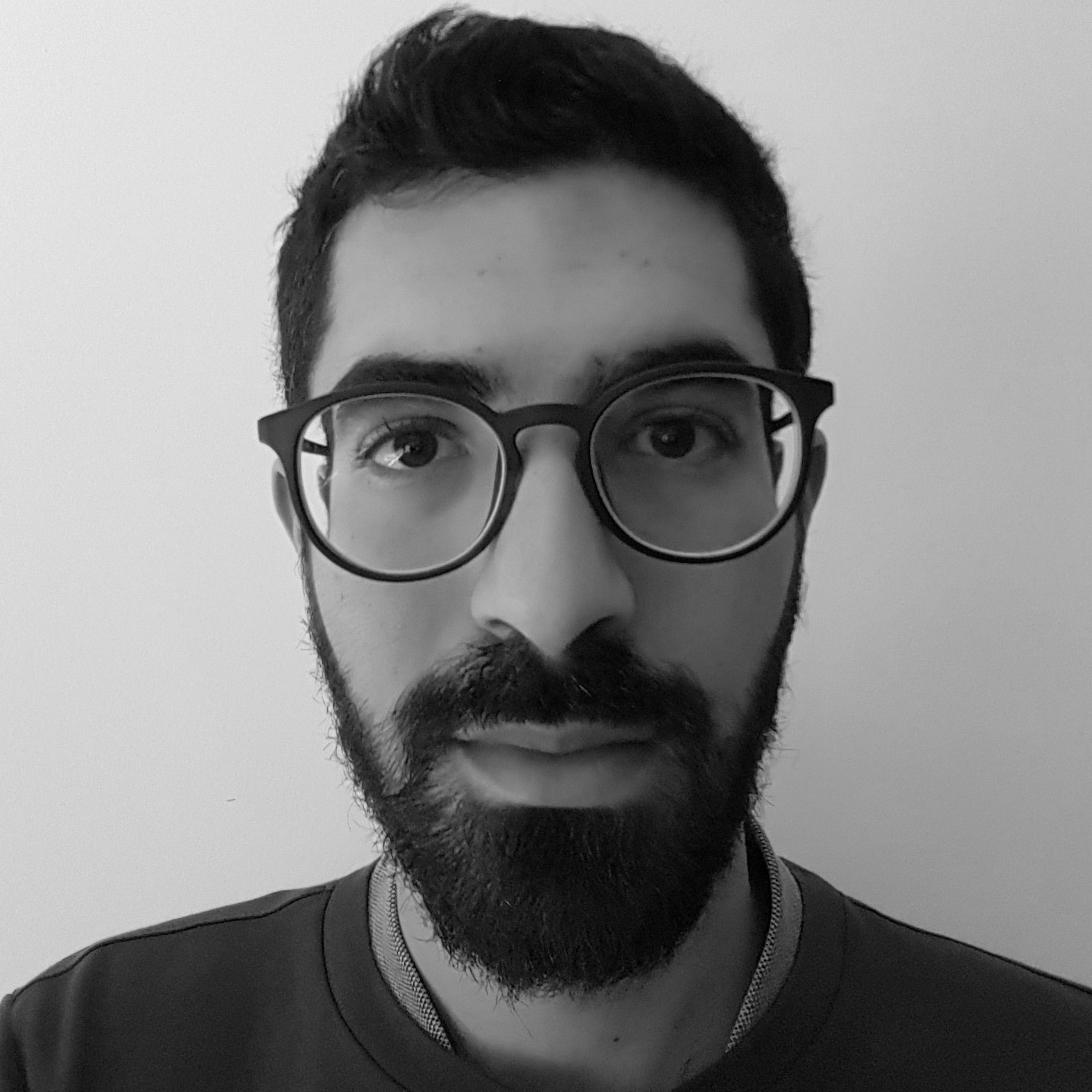}}]{Antonis Tsigkanos}{\space}(M'17) received his Ph.D., M.Sc. from the Department of Informatics and Telecommunications, of the National and Kapodistrian University of Athens (NKUA), Greece and his B.Eng. degree from the Electrical and Computer Engineering School of the National Technical University of Athens (NTUA).\par
		Currently, he works as ASIC design engineer in the semiconductor industry, designing ultra Low 
		power processors optimized for AI workloads and graphics. He has published multiple papers in peer reviewed transactions, journals, and conference proceedings. His research interests include on-board payload  data processing systems, SoC design, reliability, deep learning accelerators and low power design.
		\end{IEEEbiography}
	
		\begin{IEEEbiography}[{\includegraphics[width=1in,height=1.25in,clip,keepaspectratio]{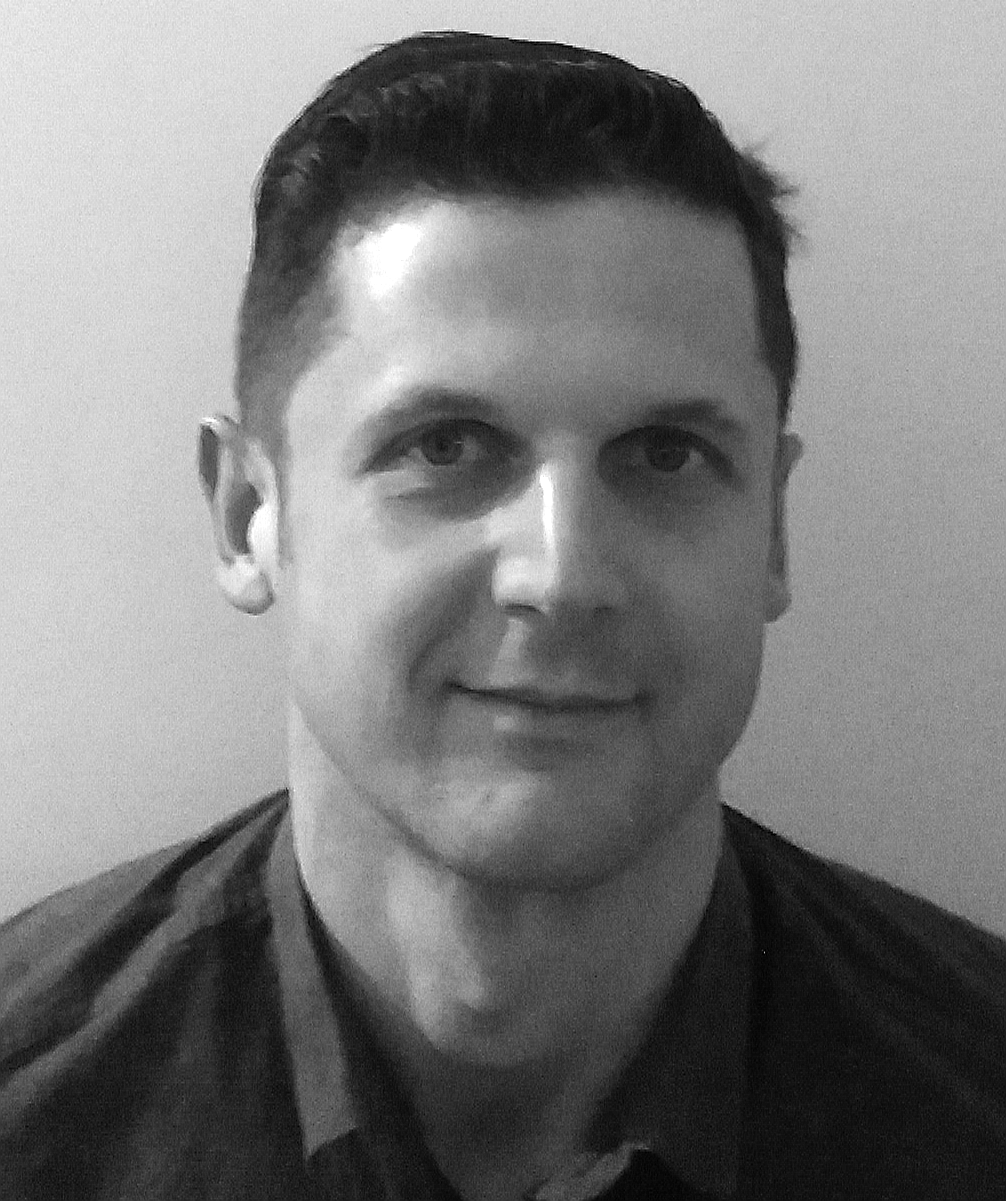}}]			{Dimitris Theodoropoulos} (M'19) received the M.Sc. degree from the Department of Informatics and Telecommunications of the National and Kapodistrian University of Athens (NKUA), Greece, Athens, Greece, in 2015.\par
		He is serving as an active duty military officer for the Hellenic Air Force (HAF), with specialty and expertise on informatics and telecommunication systems. He is pursuing his PhD degree at the the NKUA, Greece and he is a member of the Digital Systems \& Computer Architecture Laboratory (DSCAL) of NKUA. His scientific interests focus on the hardware design of onboard processing hardware systems for aerospace applications.
		\end{IEEEbiography}
	
		\vskip 0pt plus -1.1fil
		
		\begin{IEEEbiography}[{\includegraphics[width=1in,height=1.25in,clip,keepaspectratio]{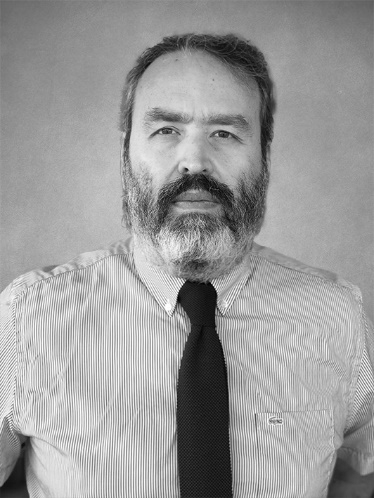}}]{Nektarios Kranitis}{\space}(Senior Member, IEEE) received the BSc degree in Physics from the Department of Physics, University of Patras, Patras, Greece, in 1997, and the PhD degree in Computer Science from the Department of Informatics and Telecommunications, of the National and Kapodistrian University of Athens (NKUA), Greece, in 2005 under Scholarship from the Institute of Informatics and Telecommunications, National Centre for Scientific Research (N.C.S.R.) “Demokritos”.\par
		He is currently an Associate Professor at the Department of Aerospace Science and Technology of NKUA, Greece. His research interests focus on on-board computers \& data handling, onboard payload data processing systems, FPGA-based acceleration and dependable and reliable systems design. He has been involved in several R\&D projects funded by ESA, EU and the Greek government as PI or senior researcher in onboard data systems technology. He has published more than 70 papers in peer reviewed transactions, journals and conference proceedings. Currently, there are more than 1500 citations that refer to his published work while his h-index is 20. He is a Senior member of the IEEE, the IEEE Aerospace and Electronic Systems Society (AESS) and the IEEE Computer Society (CS).
		\end{IEEEbiography}
	 	
	 	\vskip 0pt plus -1.11fil
	 	
	 	\begin{IEEEbiography}[{\includegraphics[width=1in,height=1.25in,clip,keepaspectratio]{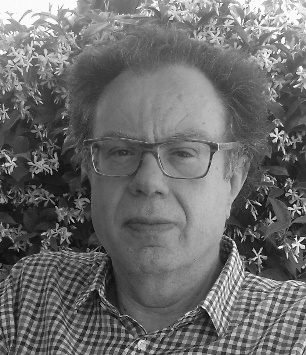}}]{Antonis Paschalis}
 		{\space}(M’97) received the B.Sc. degree in Physics, the M.Sc. degree in Electronic Automation, and the Ph.D. degree in Dependable Computers, under scholarship from NCSR “Demokritos”, all from the Department of Physics, National and Kapodistrian University of Athens (NKUA), Athens, Greece, in 1983, 1985, and 1987, respectively.\par
 		He is a Professor with the Department of Informatics and Telecommunications, NKUA, and head of space upstream technology research group of Digital Systems and Computer Architecture Laboratory participating in ESA space missions. He has published over 160 papers (26 IEEE Transactions) and holds a U.S. patent. His current research interests include reconfigurable payload data processing units and high-speed IP cores for aerospace systems, VLSI design and testing, and dependable computer architecture. Currently, there are more than 2,600 citations that refer to his published material and his h-index is 29. Prof. Paschalis is a Golden Core Member of IEEE Computer Society.
	 	\end{IEEEbiography}
		
		
		
		

	\end{document}